\documentclass[aps,prl,twocolumn,floatfix,showpacs,citeautoscript,longbibliography,hyperlinks]{revtex4-2}
\usepackage{amsmath}
\usepackage{epstopdf}
\usepackage{amsmath}
\usepackage{amssymb}
\usepackage{amsfonts}
\usepackage[colorlinks=true,citecolor=blue]{hyperref}
\usepackage{graphicx,graphics}
\usepackage{enumitem}
\setlist[enumerate]{leftmargin=6mm}
\usepackage{float}
\usepackage[caption = false]{subfig}

\newcommand{\revision}[1]{{{#1}}}
\newcommand{\revisiontwo}[1]{{{#1}}}

\begin{document}
\title{Adiabatic Cooper pair splitter}
\author{Fredrik Brange}
\affiliation{Department of Applied Physics, Aalto University, 00076 Aalto, Finland}
\author{Riya Baruah}
\affiliation{Department of Applied Physics, Aalto University, 00076 Aalto, Finland}
\author{Christian Flindt}
\affiliation{Department of Applied Physics, Aalto University, 00076 Aalto, Finland}

\begin{abstract}
Recent experiments have observed Cooper pair splitting in quantum dots coupled to superconductors, and efficient schemes for controlling and timing the splitting process are now called for. Here, we propose and analyze an adiabatic Cooper pair splitter that can produce a regular flow of spin-entangled electrons in response to a time-dependent and periodic gate voltage. The splitting process is controlled by moving \revisiontwo{adiabatically} back and forth along an avoided crossing between the empty state and the singlet state of two quantum dots that are coupled to a superconductor, followed by the emission of the split Cooper pairs into two normal-state drains. The scheme does not rely on fine-tuned resonance conditions and is therefore robust against experimental imperfections in the driving signal. We identify a range of driving frequencies, where the output currents are quantized and proportional to the driving frequency combined with suppressed low-frequency noise. We also discuss the main sources of cycle-missing events and evaluate the statistics of electrons emitted within a period of the drive as well as the distribution of waiting times between them. Realistic parameter estimates indicate that the Cooper pair splitter can be operated in the gigahertz regime.
\end{abstract}

\maketitle

\emph{Introduction.}--- Cooper pair splitters are experiencing a surge of interest as several promising experiments have brought the field closer to the ultimate goal of detecting and exploiting the non-local entanglement of split Cooper pairs~\cite{Lesovik2001,Recher2001,Hofstetter:Cooper,PhysRevLett.104.026801,PhysRevLett.107.136801,PhysRevLett.109.157002,Herrmann:Spectroscopy,Das2012,PhysRevB.90.235412,PhysRevLett.114.096602,PhysRevLett.115.227003,Borzenets:High,Bruhat2018,Baba:2018,tan2020,Ranni2021,Hofstetter:Cooper,PhysRevLett.107.136801,Herrmann:Spectroscopy,PhysRevLett.104.026801,PhysRevLett.115.227003,Borzenets:High,tan2020,Das2012,Ranni2021,Pandey:2021,Scherubl2022,Kurtossy:2022,Ranni:2022,Bordoloi:2022,Wang:2022,deJong:2022,Wang:2022b,Bordin:2023}. \revision{Since entanglement is a critical resource for most quantum applications, Cooper pair splitters have attracted attention across a wide range of fields in quantum science and technology.} Recently, Cooper pair splitting has been observed with charge detectors~\cite{Ranni2021,Ranni:2022} and dispersive read-out~\cite{deJong:2022}, correlations between spin currents have been measured~\cite{Bordoloi:2022}, and Cooper pair splitters with triplet pairing have been realized~\cite{Wang:2022,Wang:2022b}. The thermoelectric properties of Cooper pairs splitter have also been explored in theory~\cite{Cao2015,Sanchez2018,Hussein2019,Kirsanov2019} and experiment~\cite{tan2020}. Cooper pair splitters have been implemented in a variety of  architectures based on  nanowires~\cite{Hofstetter:Cooper,PhysRevLett.107.136801,Das2012,PhysRevB.90.235412,PhysRevLett.115.227003,Baba:2018,Scherubl2022,Kurtossy:2022,Bordoloi:2022,Wang:2022,deJong:2022}, carbon nanotubes~\cite{PhysRevLett.104.026801,PhysRevLett.109.157002,Herrmann:Spectroscopy,Bruhat2018}, graphene~\cite{PhysRevLett.114.096602,Borzenets:High,tan2020,Pandey:2021}, semiconductor quantum dots~\cite{deJong:2022}, and metallic islands~\cite{Ranni2021,Ranni:2022}. Very recently, setups with several quantum dots and superconductors have also been experimentally realized~\cite{Bordin:2023}.

These experimental advances have reduced the gap between experiment and theory, and several theoretical ideas for future experiments may soon be within reach. As an example, the distribution of waiting times was already measured following a recent suggestion~\cite{Walldorf2018,Ranni2021}. There are also proposals for observing the entanglement of the split Cooper pairs by either violating a Bell inequality \cite{Kawabata2001,Sauret:Spin,Braunecker2013,Busz:2017} or by using an entanglement witness formulated in terms of cross-correlation measurements of the outgoing spin currents~\cite{Klobus:2014,Brange2017,Tam2021}. In addition, several Cooper pair splitters may be combined to create a Kitaev chain with Majorana bound states forming at the ends~\cite{Leijnse:2012,Sau:2012,Fulga:2013,Wang:2022,Bordin:2023}. Moreover, while experiments have focused on static devices, there are also proposals to control the splitting of Cooper pairs using time-dependent drives~\cite{Hiltscher2011,Brange:2021}. In this context, it is an open question how one should design the driving scheme in the best way.

\begin{figure}[b!]
	\centering
	\includegraphics[width=0.94\columnwidth]{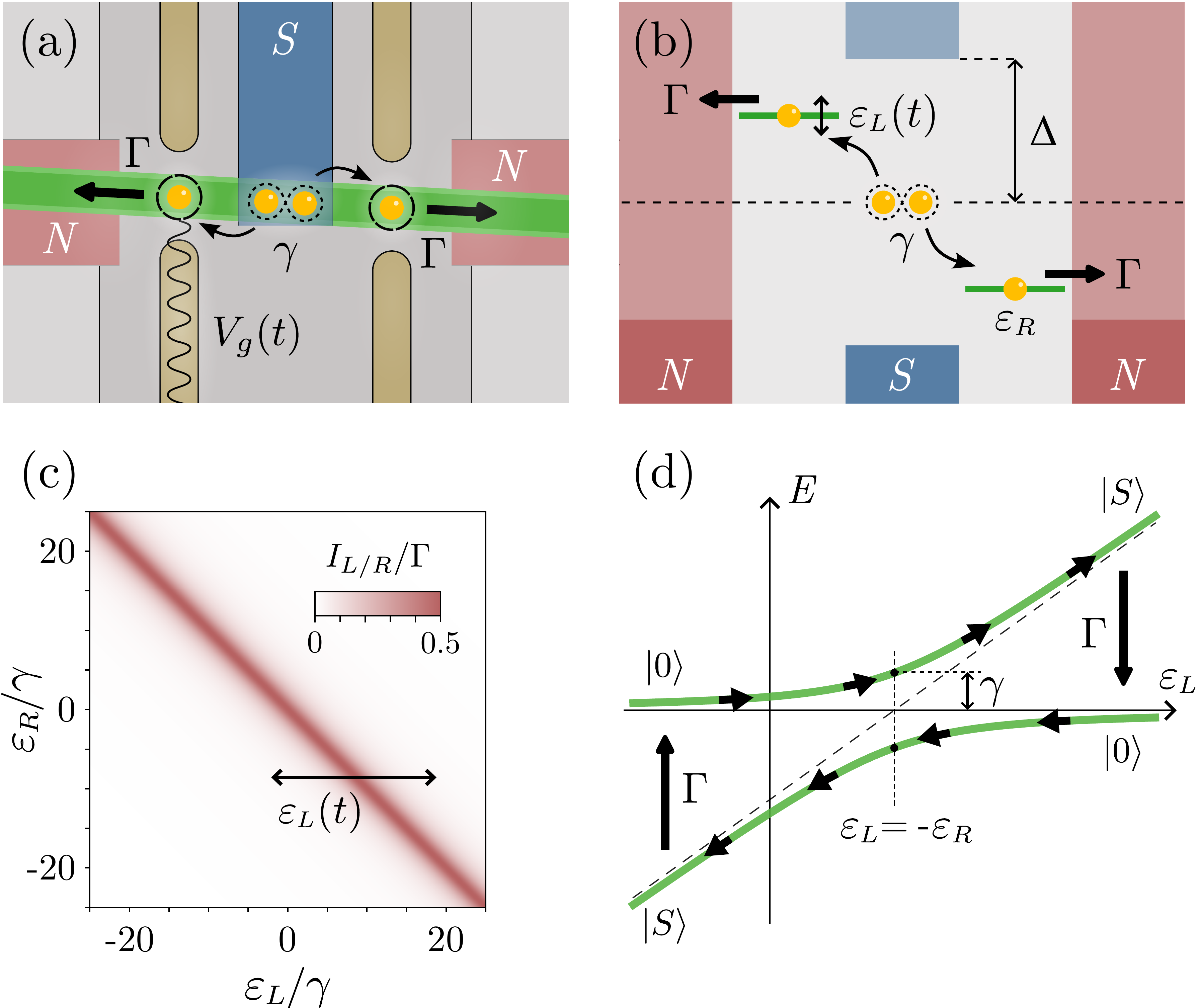}
	\captionsetup{justification=justified,singlelinecheck=false}
	\caption{Adiabatic Cooper pair splitter. (a) The device consists of a nanowire \revision{(green)} with gate-defined quantum dots coupled to a superconductor ($S$\revision{, blue}). The amplitude for Cooper pair splitting is denoted by $\gamma$, while  $\Gamma$ is the rate at which electrons are emitted into the normal-state electrodes ($N$\revision{, red}). A time-dependent gate voltage, $V_g(t)$, is used to control the left quantum dot level. (b) The superconducting gap is denoted by $\Delta$, and $\varepsilon_{L/R}$ are the tunable level positions. (c) Current as a function of the level positions for a static device with $\hbar\Gamma/\gamma=0.01$. In our scheme, we move the left level back and forth across the peak in the current. (d) \revision{The amplitude for Cooper pair splitting leads to an avoided crossing between the singlet state~$|S\rangle$ and the empty state~$|0\rangle$.  We move back and forth along the avoided crossing, and a split Cooper pair is emitted into the drains after each crossing.}}
	\label{Cooper pair splitter}
\end{figure}

\begin{figure*}
	\centering
	\includegraphics[width=\textwidth]{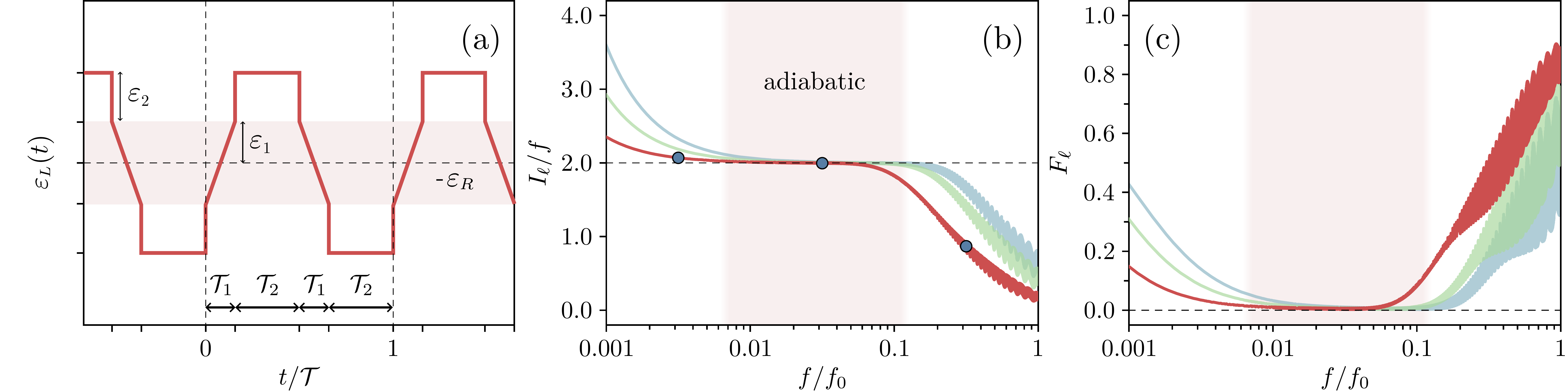}
	\captionsetup{justification=justified,singlelinecheck=false}
	\caption{Adiabatic driving scheme, average current, and low-frequency noise. (a) The period of the drive is divided into two splitting phases and two emissions phases, each of duration $\mathcal{T}_1$ and $\mathcal{T}_2$, respectively, such that $\mathcal{T}=2(\mathcal{T}_1+\mathcal{T}_2)$. During the splitting phases, the left level is moved across the resonance, $\varepsilon_L=-\varepsilon_R$, and a Cooper pair is split. During the emission phases, the levels are kept far off resonance, and the split Cooper pair tunnels into the drains. (b) Average current as a function of the driving frequency. The parameters are $\kappa=\gamma$, $\varepsilon_1=50\gamma$, $\varepsilon_2=100\gamma$,  $\varepsilon_R=-100\gamma$, and $\hbar\Gamma=0.001\gamma$ (red), $0.002\gamma$ (green), and $0.003\gamma$ (blue), and we have defined $f_0=\alpha\gamma/2\pi\hbar$ with $\alpha=\mathcal{T}_1/\mathcal{T}=0.01$. The adiabatic regime, where the current should taken on the value $I_\ell = 2f$, is indicated for the red curve by the shaded area according to Eq.~(\ref{eq:require}). (c) The Fano factor, $F_\ell=S_\ell/I_\ell$, as a function of the driving frequency. The three circles in panel (b) indicate the frequencies used in Figs.~\ref{Fig3}~and~\ref{Fig4}. }
	\label{Fig2}
\end{figure*}

In this Letter, we propose and analyze an adiabatic Cooper pair splitter that operates by driving two quantum dots coupled to a superconductor back and forth along an avoided crossing between the empty state and the singlet state of the quantum dots, see Fig.~\ref{Cooper pair splitter}. Each time the dots are filled by a split Cooper pair from the superconductor, the system is taken back to the empty state as the electrons are emitted into the drain electrodes, and the process can repeat. The avoided crossing occurs because of the \revision{amplitude of Cooper pair splitting, which couples the singlet state of the quantum dots with the empty state}, and the driving can be implemented with an external gate. When operated adiabatically, the Cooper pair splitter delivers a regular and low-noise flow of split Cooper pairs as shown in Fig.~\ref{Fig2}. The scheme does not rely on fine-tuned resonance conditions or accurate timing and may be realized based on recent experiments.

\emph{Adiabatic Cooper pair splitter.}--- Figure~\ref{Cooper pair splitter}(a) shows the Cooper pair splitter consisting of two single-level quantum dots coupled to a superconductor. We here consider a setup based on quantum dots along a nanowire, \revision{similar to recent experiments \cite{Hofstetter:Cooper,Kurtossy:2022,Bordoloi:2022,Wang:2022,deJong:2022}}, but our proposal would also work for other architectures. With a large superconducting gap, the dynamics of the quantum dots can be described by the effective Hamiltonian~\cite{Eldridge:Superconducting,Sauret:Quantum,Hiltscher2011,Walldorf2020,Brange:2021} 
\begin{equation}\label{eq:Heff}
\hat{H}=\sum_{\ell\sigma}\varepsilon_\ell\hat{d}_{\ell\sigma}^\dagger \hat{d}_{\ell\sigma}^{\phantom\dagger}-\gamma(\hat d_S^\dagger+\hat d_S) -\kappa\sum_\sigma (\hat{d}_{L\sigma}^\dagger \hat{d}_{R\sigma}^{\phantom\dagger}+
\mathrm{h.c.}),
\end{equation}
where $\varepsilon_\ell$ are the energy levels of the quantum dots,  $\ell =L,R$, which can be tuned by external gates to 
control the splitting of Cooper pairs. The amplitudes for Cooper pair splitting and elastic cotunneling are denoted by $\gamma$ and $\kappa$, respectively. The operator $\hat d_{\ell\sigma}^\dagger$ creates electrons with spin $\sigma =\, \uparrow,\downarrow$ in either of the dots, while $\hat d_S^\dagger \equiv (\hat{d}_{L\downarrow}^\dagger \hat{d}_{R\uparrow}^\dagger-\hat{d}_{L\uparrow}^\dagger \hat{d}_{R\downarrow}^\dagger)/\sqrt{2}$ describes a singlet state that is delocalized between them. Strong Coulomb interactions on the quantum dots prevent each of them from being doubly occupied, which ensures that the electrons from a split Cooper pair tunnel into different quantum dots. We work with a large detuning of the dot levels, $|\varepsilon_L-\varepsilon_R| \gg \kappa$, to suppress elastic cotunneling between them. The empty state of the quantum dots, $|0\rangle$, with zero energy is coherently coupled to the singlet state, $|S\rangle=\hat d_S^\dagger|0\rangle$, with energy $\varepsilon_L(t)+\varepsilon_R$ by the amplitude for Cooper pair splitting, $\gamma$, and the energy of the singlet state is controlled by a time-dependent gate voltage on the left quantum dot. 

\begin{figure*}
	\centering
	\includegraphics[width=\textwidth]{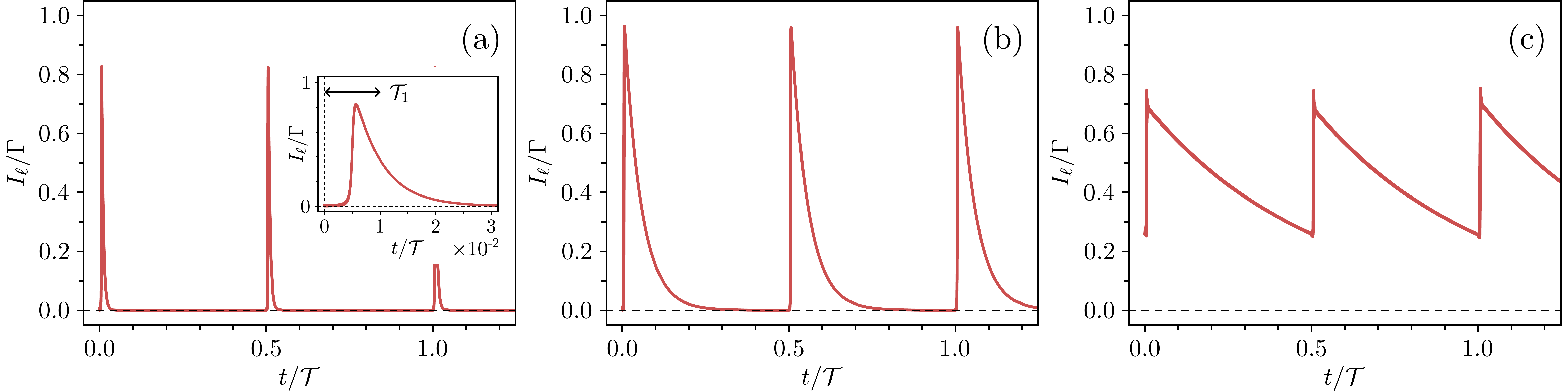}
	\captionsetup{justification=justified,singlelinecheck=false}
	\caption{Time-dependent currents. We show the time-dependent currents corresponding to the three points marked with circles in Fig.~\ref{Fig2}. (a) At low frequencies, more than one electron is emitted per half-period, and emissions occur already in the splitting phase, see inset. (b) In the adiabatic regime, one electron is emitted at every half-period, and the leakage current in the splitting phase is suppressed. (c) At high frequencies, the regularity is gradually lost, and there is always a finite current running.}
	\label{Fig3}
\end{figure*}

As shown in Fig.~\ref{Cooper pair splitter}(b), large voltages are applied to the normal-state electrodes, so that they function as drains for the dots. Without a time-dependent drive, the (particle) currents running into the drains are
\begin{equation}
	I_{L/R} =\frac{2\Gamma \gamma^2}{(\varepsilon_L+\varepsilon_R)^2+(\hbar\Gamma)^2 +4\gamma^2}, 
	\label{eq:static_curr}
\end{equation}
where $\Gamma$ is the tunneling rate into the drains, which we assume to be the same for the two drains to keep the discussion simple~\cite{Note1,Sauret:Quantum,Walldorf2020}. In Fig.~\ref{Cooper pair splitter}(c), we show the current as a function of the level positions, and we see a peak along the diagonal $\varepsilon_R=-\varepsilon_L$, where the singlet state is on resonance with the empty state. A similar dependence was observed in the recent experiments of Refs.~\cite{Wang:2022,Wang:2022b}.

\emph{Driving scheme.}---  To describe the adiabatic driving scheme, we show in Fig.~\ref{Cooper pair splitter}(d) the energy of the empty state and the singlet state as a function of $\varepsilon_L$, and we observe an avoided crossing  between them at $\varepsilon_L=-\varepsilon_R$ because of the coupling $\gamma$.  Thus, if we start with a large value of $\varepsilon_L$, the empty state will have the lowest energy, and as we move through the avoided crossing by decreasing $\varepsilon_L$, the quantum dots will eventually become occupied by a split Cooper pair. At the same time, the probability increases for the electrons to leave the dots via the drains. The system thereby returns to the empty state, which now has a higher energy than the singlet state. After that, we increase the energy of the \revision{left level}, and we again move from the empty state to the singlet state, but this time following the excited state of the system. Eventually, the quantum dots are again occupied by a split Cooper pair, and once again the system is taken back to the empty state as the electrons tunnel into the drains. By doing so periodically, two split Cooper pairs should be produced per period of the drive.

The driving scheme in Fig.~\ref{Fig2}(a) is now designed with the following requirements in mind. To formulate them, we divide the period of the drive into four phases, two splitting phases, each of duration $\mathcal{T}_1$, and two emission phases, each of duration $\mathcal{T}_2$. The period of the drive is then $\mathcal{T} =2(\mathcal{T}_1+\mathcal{T}_2)=1/f$, where $f$ is the driving frequency. Our requirements for the drive are now: 
\begin{enumerate}
	\item \emph{Adiabatic splitting:} The splitting phase should start off resonance, so that $\gamma/\varepsilon_1\ll 1$, and the drive should be slow, so that $\gamma T_1/\hbar\times \gamma/\varepsilon_1\gg 1$, where $\varepsilon_1=\varepsilon_L+\varepsilon_R$ is the singlet energy at the onset~\cite{Shevchenko:2010,Ivakhnenko:2023}. 
	\item \emph{No leakage:} To make sure that no electrons are  emitted during the splitting phase, we need~$\Gamma\mathcal{T}_1\ll 1$.
	\item \emph{Emission:} To ensure that the electrons are emitted during the emission phase, we need~$\Gamma\mathcal{T}_2\gg 1$. Also, during the emission phase, Cooper pair splitting should be off resonance, so that $|\varepsilon_L+\varepsilon_R|\gg\gamma$.
\end{enumerate}
These requirements can be combined into the inequality
\begin{equation}
\alpha	\Gamma\ll f\ll \mathrm{min}\{\alpha\gamma^2/\hbar\varepsilon_1,\Gamma/2\},
	\label{eq:require}
\end{equation}
which specifies the range of possible driving frequencies for the adiabatic Cooper pair splitter given a fixed ratio of the splitting time over the period of the drive, $\alpha=\mathcal{T}_1/\mathcal{T}$. To provide realistic estimates, we note that the amplitude for Cooper pair splitting can be on the order of $\gamma= 40$ $\mu$eV together with tunneling rates of $\hbar\Gamma = 4$ $\mu$eV (or 1 GHz)~\cite{Hofstetter:Cooper,PhysRevLett.104.026801,PhysRevLett.107.136801,PhysRevLett.109.157002,Herrmann:Spectroscopy,Das2012,PhysRevB.90.235412,PhysRevLett.114.096602,PhysRevLett.115.227003,Borzenets:High,Bruhat2018,Baba:2018,tan2020,Ranni2021,Hofstetter:Cooper,PhysRevLett.107.136801,Herrmann:Spectroscopy,PhysRevLett.104.026801,PhysRevLett.115.227003,Borzenets:High,tan2020,Das2012,Ranni2021,Pandey:2021,Scherubl2022,Kurtossy:2022,Ranni:2022,Bordoloi:2022,Wang:2022,deJong:2022,Wang:2022b,Bordin:2023}. Taking the duration of the splitting phase so that $\alpha=0.1$, combined with a singlet energy of $\varepsilon_1=100$ $\mu$eV at the onset, the inequality~(\ref{eq:require}) predicts adiabatic frequencies in the range 50 MHz $\ll f\ll $ 500 MHz. For example, with a driving frequency of $f=100$ MHz, we would expect currents of about 20 pA, since two electrons are emitted into each drain per period of the drive. We may also take $\varepsilon_2=100$ $\mu$eV in Fig.~\ref{Fig2}(a) to suppress Cooper pair splitting during the emission phase.
 
\emph{Average current.}---  To illustrate the operation of the Cooper pair splitter, we calculate the drain currents. To this end, we consider the density matrix of the dots, $\hat \rho(t)$, whose dynamics obeys the Lindblad equation~\cite{Sauret:Quantum,Walldorf2020,PhysRevB.63.165313}
\begin{equation}\label{eq:vonNeumannEq}
\frac{d}{dt}\hat \rho(t)=\mathcal{L}(t)\hat \rho(t)=\frac{1}{i\hbar}[\hat H(t),\hat \rho(t)]+\mathcal{D}\hat \rho(t).
\end{equation}
Here, tunneling  to the drains is described by the term
\begin{equation}
\mathcal{D}\hat \rho(t)=\Gamma\sum_{\ell\sigma}\big( \hat d_{\ell\sigma}^{\phantom\dagger}\hat \rho(t) \hat d_{\ell\sigma}^\dagger-\frac{1}{2}\{\hat \rho(t),\hat d_{\ell\sigma}^\dagger \hat d_{\ell\sigma}^{\phantom\dagger}\}\big),
\label{eq:Dissipator}
\end{equation}
and the Hamiltonian $\hat H(t)$ is given by Eq.~(\ref{eq:Heff}) with time-dependent levels. \revisiontwo{Because of the large voltages, the chemical potential of the drains is placed well below the quantum dot levels, so that the thermal smearing due to a finite temperature becomes unimportant.} Single-electron excitations above the gap are exponentially suppressed in the ratio of the superconducting gap over the temperature as $\exp(-\Delta/k_BT)$, allowing us to ignore such excitations. Realistically, the gap can be up to $\Delta \simeq 1 $ meV  (corresponding to a temperature of about 10 K), which indeed is much higher than typical experimental temperatures of around $T=100$ mK.

\begin{figure*}
	\centering
	\includegraphics[width=\textwidth]{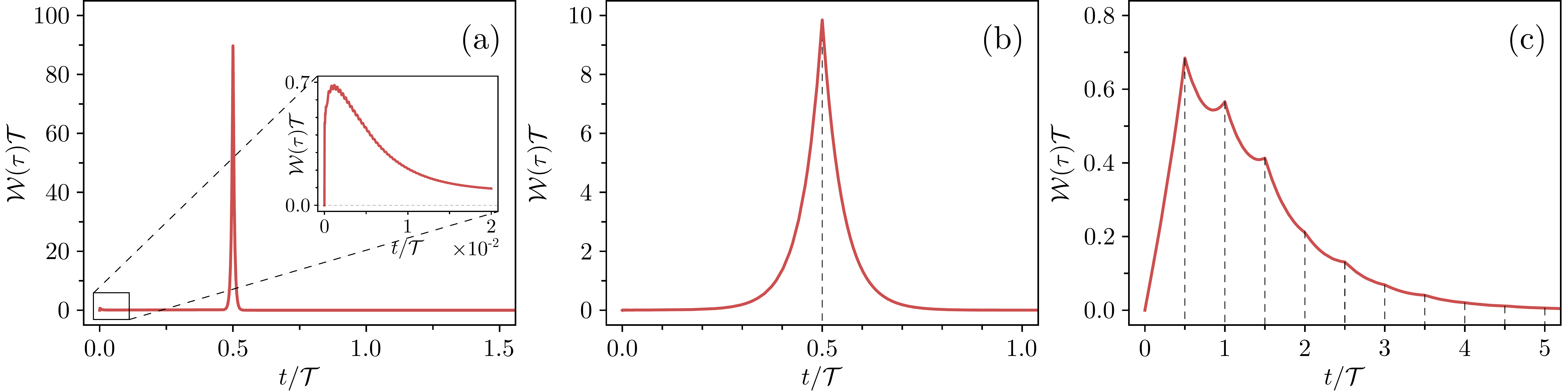}
	\captionsetup{justification=justified,singlelinecheck=false}
	\caption{Distribution of waiting times.  We show distributions corresponding to the three points marked with circles in Fig.~\ref{Fig2}. (a) At low frequencies, a peak develops at short waiting times. (b) In the adiabatic regime, a single peak at half the period shows that Cooper pairs are being split periodically. (c) At high frequencies, cycle-missing events give rise to several peaks.}
	\label{Fig4}
\end{figure*}

Figure~\ref{Fig2}(b) shows the current as a function of the driving frequency~\cite{Note1}. When operated in the adiabatic regime, the device should deliver two split Cooper pairs per period of the drive, and the drain currents should take on the quantized value $I_\ell=2f$. This expectation is confirmed by our calculations, which show a quantized current in the adiabatic regime defined by Eq.~(\ref{eq:require}). At higher frequencies, the number of emitted electrons drops off, since the driving becomes too fast, and a Cooper pair is not split in each crossing of the resonance. The current does not vanish at low frequencies, since the system is biased, and a current will run even without the drive. \revision{For this reason, the ratio of the current over the frequency can become large at very low frequencies.} Experimentally, the plateau in Fig.~\ref{Fig2}(b) would demonstrate the adiabatic splitting of Cooper pairs.

\emph{Noise and Fano factor.}---  To further analyze the splitting of Cooper pairs, we show in Fig.~\ref{Fig2}(c) the low-frequency noise $S_\ell$ of the drain currents, quantified by the Fano factor, $F_\ell=S_\ell/I_\ell$~\cite{Note1,Blanter:2000,Bagrets:2003,Pistolesi:2004,Flindt2005,Benito:2016,Potanina:2019}. In the adiabatic regime, we expect a strong suppression of the noise, which indeed is confirmed by our calculations. By contrast, at lower frequencies, the Fano factor increases and comes closer to the values for a static device~\cite{Walldorf2020}. At high frequencies, the splitting of Cooper pairs becomes rare and uncorrelated, and the Fano factor approaches one. We also observe oscillations in the current and the Fano factor, which can be attributed to an interplay between the amplitude of Cooper pair splitting and the driving frequency. However, for our purposes, we focus on the noise in the adiabatic regime, which provides another experimental signature of the regular splitting of Cooper pairs. Unlike the current, which should be measured over a range of frequencies to observe the plateau in Fig.~\ref{Fig2}(b), the low noise can be measured at just a single frequency.

\emph{Cycle-missing events.}--- As the driving frequency is increased beyond the adiabatic regime, we expect cycle-missing events to occur because of non-adiabatic excitations~\cite{Shevchenko:2010,Ivakhnenko:2023}. In particular, the system may make transitions between the instantaneous eigenstates, if we move too fast along the avoided crossing. Also, if the unloading phase is too short compared with the escape time to the drains, a split Cooper pair may not be emitted into the drains, and it might be transferred back into the superconductor. If we denote the small probability of a cycle-missing event by $p\ll 1$, the current will be reduced to $I_\ell=2f(1-p)$, while the noise increases from zero to $S_\ell=2f p$~\cite{Albert2010}. The Fano factor can then be approximated as $F_\ell\simeq p$, showing that it directly measures the probability of cycle-missing events. In Fig.~\ref{Fig2}(c),  the Fano factor becomes as small as one percent, noting that we are aiming for a periodic emitter of entangled electrons rather than metrological applications, which often require error rates below parts per million~\cite{Pekola:2013}.

\emph{Time-dependent current.}--- It is instructive also to consider the time-dependent currents $I_\ell(t)$, which provide information about the statistics of electrons emitted within a period of the drive. In Fig.~\ref{Fig3}, we show the time-dependent currents for the three points marked with circles in Fig.~\ref{Fig2}. At low frequencies, we enter the quasi-static regime, where the current approaches the static result in Eq.~(\ref{eq:static_curr}) with the time-dependent level position inserted. By contrast, in the adiabatic regime, the time-dependent current shows how the quantum dots are periodically filled by a split Cooper pair, followed by the emission of the electrons into the drains. At higher frequencies, the driving becomes non-adiabatic, such that the quantum dots are not filled or emptied in every half-period, and there is always a finite current running.

\emph{Distribution of waiting times.}--- Finally, we turn to the distribution of electron waiting times~\cite{Brandes:Waiting,Albert2011}, which were recently measured for a static Cooper pair splitter~\cite{Walldorf2018,Ranni2021}. Here, we consider the distribution of the time that passes between electrons tunneling into one of the drains~\cite{Brandes:Waiting,Walldorf2018,Brange:2021,Note1}. In Fig.~\ref{Fig4}, we show distributions for the three points marked with circles in Fig.~\ref{Fig2}. At low frequencies, a peak develops at short times, corresponding to several emissions occurring as the current resonance is crossed. On the other hand, in the adiabatic regime, a single peak at half the period shows that Cooper pairs are split periodically, with the width of the peak given by the tunneling rate to the drains. Finally, in the non-adiabatic regime, peaks appear at multiples of the half-period, since cycle-missing events start to occur.

\emph{Conclusions.}--- We have proposed and analyzed an adiabatic Cooper pair splitter that operates by moving a quantum dot level back and forth along an avoided crossing. Each time the resonance is crossed, a Cooper pair is split and emitted from the quantum dots. When operated in the adiabatic regime, the device generates a regular flow of spin-entangled electrons with currents that are proportional to the driving frequency combined with vanishing low-frequency noise. Our proposal appears feasible in the light of recent experiments, and it can be extended in many directions.  For example, it may be possible to increase the driving frequency with a shortcut to adiabaticity~\cite{Odelin:2019}. Moreover, in materials like InAs,  one may use the spin-orbit coupling combined with time-dependent gates to rotate the spins in the dots~\cite{Flindt:2006,Golovach:2006}. One may also envision Cooper pair splitters that are coupled to ballistic conductors so that the entangled electrons can be transferred to other parts of a solid-state circuit for further operations, manipulation, and read-out.

\acknowledgements
\emph{Acknowledgements.}---  We acknowledge support from the Nokia Industrial Doctoral School in Quantum Technology and the Research Council of Finland through the Finnish Centre of Excellence in Quantum Technology (grant number 352925) and grant number 331737.

\end{document}


\title{Supplemental Material for ``Adiabatic Cooper Pair Splitter"}
\author{Fredrik Brange}
\affiliation{Department of Applied Physics, Aalto University, 00076 Aalto, Finland}
\author{Riya Baruah}
\affiliation{Department of Applied Physics, Aalto University, 00076 Aalto, Finland}
\author{Christian Flindt}
\affiliation{Department of Applied Physics, Aalto University, 00076 Aalto, Finland}

\maketitle

\onecolumngrid

\renewcommand{\theequation}{S\arabic{equation}}%
\renewcommand{\thefigure}{S\arabic{figure}}%

\section{Lindblad equation \& vectorization}\label{sec:mastereq}
\noindent
As explained in the main text, the Cooper pair splitter can be described by the Lindblad equation~\cite{Sauret:Quantum,Walldorf2020}
\begin{equation}\label{eq:vonNeumannEq}
	\frac{d}{dt}\hat \rho(t)=\mathcal{L}(t)\hat \rho(t)=\frac{1}{i\hbar}[\hat H(t),\hat \rho(t)]+\Gamma\sum_{\ell\sigma}\big( \hat d_{\ell\sigma}^{\phantom\dagger}\hat \rho(t) \hat d_{\ell\sigma}^\dagger-\frac{1}{2}\{\hat \rho(t),\hat d_{\ell\sigma}^\dagger \hat d_{\ell\sigma}^{\phantom\dagger}\}\big),
\end{equation}
where 
\begin{equation}\label{eq:Heff}
	\hat{H}(t)=\sum_{\ell\sigma}\varepsilon_\ell(t)\hat{d}_{\ell\sigma}^\dagger \hat{d}_{\ell\sigma}^{\phantom\dagger}-\gamma(\hat d_S^\dagger+\hat d_S) -\kappa\sum_\sigma (\hat{d}_{L\sigma}^\dagger \hat{d}_{R\sigma}^{\phantom\dagger}+
	\mathrm{h.c.}),
\end{equation}
is the effective Hamiltonian of the two quantum dots with time-dependent level positions, $\varepsilon_\ell(t)$. To carry out our calculations, we vectorize the density matrix of the quantum dots and implement a matrix representation of the Liouvillian. Here, we do not explicitly consider the spin-degrees of freedom, and it therefore suffices to express the density matrix and the Liouvillian in the charge basis only. In this representation, the density matrix takes the form
\begin{equation}
	\hat\rho= \left( 
	\begin{array}{cccc}
		\rho_{00} & 0 & 0 &  \rho_{S0} \\
		0 & \rho_{LL}  & \rho_{RL} & 0  \\
		0 & \rho_{LR} & \rho_{RR} & 0\\
		\rho_{0S} & 0 & 0 & \rho_{SS}
	\end{array}
	\right),
\end{equation}
where $\rho_{\ell \ell'}=\sum_{\sigma}\rho_{\ell\sigma,\ell'\sigma}$ are given by traces over the spins. The density matrix can be written on the vectorized form 
\begin{equation}
	\hat\rho= (\rho_{00},\rho_{LL},\rho_{RR},\rho_{SS},\rho_{0S}, \rho_{S0},\rho_{LR},\rho_{RL})^T, 
\end{equation}
where the first four elements are the populations, and the others are the coherences. The Liouvillian then becomes
\begin{equation}\label{eq:kernelmatrix}
\mathcal{L}(\chi,t)=
\left( 
	\begin{array}{cccccccc}
	0 & \Gamma e^{i\chi} & \Gamma & 0 & -i\gamma & i\gamma & 0 & 0 \\
	0 & -\Gamma  & 0 & \Gamma  & 0 & 0 &  -i\kappa &  i\kappa \\
	0 & 0 & -\Gamma  & \Gamma  e^{i\chi} & 0 & 0 & i\kappa & -i\kappa \\
	0 & 0 & 0 & -2\Gamma & i\gamma & -i\gamma & 0 & 0 \\
	-i\gamma & 0 & 0 & i\gamma & i\epsilon(t)-\Gamma  & 0 & 0 & 0 \\
	i\gamma  & 0 & 0 & -i\gamma  & 0 & -i\epsilon(t)-\Gamma  & 0 & 0 \\
	0 & -i\kappa & i\kappa & 0 & 0 & 0 & -i\delta(t)-\Gamma  & 0 \\
	0 & i\kappa & -i\kappa & 0 & 0 & 0 & 0 & i\delta(t)-\Gamma 
	\end{array}
\right),
\end{equation}
where we have included a counting field, $\chi$, that couples to transitions into the left lead, and we have defined the detuning and the sum of the energy levels, $\delta=\varepsilon_L-\varepsilon_R$ and $\epsilon=\varepsilon_L+\varepsilon_R$, with $\hbar,e=1$ from now on. We note that a matrix representation of the spin-resolved Liouvillian can be found in the appendix of Ref.~\cite{Walldorf2020}.

\section{Time-dependent \& period-averaged current}\label{sec:current}
\noindent
In the main text, we show results for the time-dependent current running into the left lead, given by the expression
\begin{equation}
	I_L(t)=\mathrm{tr}\{\mathcal{J}_{L}\hat \rho_C(t)\}
	\label{eq:time_curr}
\end{equation}
in terms of the jump operator $\mathcal{J}_{L}\hat \rho \equiv \Gamma \sum_\sigma \hat d_{L\sigma}^{\phantom\dagger}\hat \rho \hat d_{L\sigma}^\dagger$ and the periodic state of the system with the property $\hat \rho_C(t)=\hat \rho_C(t+\mathcal T)$.   To find the periodic state, we need the time-evolution operator
\begin{equation}
	\mathcal{U}(t,t_0) = \hat{T}\left\{ e^{\int_{t_0}^t \mathcal{L}(t') dt'}\right\}\simeq \prod_i e^{\mathcal{L}(t_i) \Delta t},
\end{equation} 
where $\hat{T}$ is the time-ordering operator, and $\mathcal{L}(t)$ is the Liouvillian without the counting field. We also show how we evaluate the time-evolution operator by discretizing the interval $[t_0,t]$ in small steps of size $\Delta t$, during which the Liouvillian is roughly constant. The periodic state can be found from the eigenproblem, $\mathcal{U}(t+\mathcal{T},t)\hat \rho_C(t)=\hat \rho_C(t)$, and the trace operation in Eq.~(\ref{eq:time_curr}) is implemented by summing over the first four elements of $\mathcal{J}_{L}\hat \rho_C(t)$ in its vectorized form. Moreover, the period-averaged current can be obtained as $I_L = \int_0^\mathcal{T}dt I_L(t)/\mathcal{T}$. Without the driving, one can analytically find the stationary state, defined by  $\mathcal{L}\hat\rho_S=0$, and Eq.~(2) of the main text then follows as $I_L=\mathrm{tr}\{\mathcal{J}_{L}\hat \rho_S\}$.

\section{Low-frequency noise}\label{sec:noise}
\noindent We find the low-frequency noise using techniques from full counting statistics by including a counting field as in Eq.~(\ref{eq:kernelmatrix})~\cite{Bagrets:2003,Pistolesi:2004,Flindt2005,Benito:2016,Potanina:2019}. The moment generating function for the number of emitted electrons after $N$ periods then reads 
\begin{equation}
	M(\chi,N) = \text{tr}\left\{ [\mathcal{U}(\chi,\mathcal T,0)]^N \hat\rho_C(0) \right\},
	\label{MGF}
\end{equation}
where we have defined $\mathcal{U}(\chi,t,t_0) = \hat{T} \{e^{\int_{t_0}^t \mathcal{L}(\chi,t') dt'}\}$. The cumulant generating function of the current is given as
\begin{equation}
F(\chi)=\lim_{N\rightarrow \infty}\ln[M(\chi,N)]/N\mathcal T = \ln[\max_i\{\lambda_i(\chi)\}]/\mathcal T
\end{equation}
in terms of the eigenvalue of $\mathcal{U}(\chi,\mathcal{T},0)$ with the largest absolute value. Moreover, the zero-frequency cumulants of the current are given by derivatives with respect to the counting field as $\langle \!\langle I_L^n \rangle\!\rangle = \partial_{i \chi}^n F(\chi)|_{\chi=0}$. Specifically, the average current and the noise are the first and second cumulants of the current, $I_L = \partial_{i \chi} F(\chi)|_{\chi=0}$
and $S_L = \partial^2_{i \chi} F(\chi)|_{\chi=0}$.

\section{Waiting time distribution}\label{sec:wtd}
\noindent
To evaluate the distribution of waiting times between electrons tunneling into the left lead, we use the expression
\begin{equation}
	\mathcal{W}_{L}(\tau) =\overline{\mathrm{tr}\{ \mathcal{J}_{L} \mathcal{U}_L(t+\tau,t)\mathcal{J}_{L} \hat{\rho}_C(t)\}}/ I_{L},
\end{equation} 
where the overline denotes an average over a period of the drive, while $\mathcal{U}_L(t,t_0) = \hat{T}\{e^{\int_{t_0}^t \left(\mathcal{L}(t')-\mathcal{J}_L\right) dt'} \}$ is the time-evolution operator, which excludes electron tunneling into the left drain~\cite{Brandes:Waiting,Walldorf2018,Brange:2021}. By evaluating this expression, we obtain the waiting time distributions presented in the main text.